\documentclass[journal=nalefd, layout=twocolumn]{achemso}

\usepackage[T1]{fontenc} % Use modern font encodings
\usepackage[utf8]{inputenc}
\usepackage[english]{babel}
\usepackage{amsmath,amssymb}

\usepackage{xcolor}

\author{François Rivière}
\author{Timothée de Guillebon}
\author{Damien Raynal}
\author{Martin Schmidt}
\author{Jean-Sébastien Lauret}
\author{Jean-François Roch}
\author{Loïc Rondin}
\affiliation[Paris Saclay University]
{Université Paris-Saclay, ENS Paris-Saclay, CNRS, CentraleSupélec, LuMIn, 91190 Gif-sur-Yvette, France}
\email{loic.rondin@universite-paris-saclay.fr}

\title{Hot Brownian motion of optically levitated nanodiamonds}

\begin{document}

%\begin{tocentry}
%        \includegraphics{toc_v5b.pdf}
%\end{tocentry}

\begin{abstract}
      The Brownian motion of a particle hotter than its environment is an iconic out-of-equilibrium system. Its study provides valuable insights into nanoscale thermal effects. Notably, it supplies an excellent diagnosis of thermal effects in optically levitated particles, a promising platform for force sensing and quantum physics tests. Thus, understanding the relevant parameters in this effect is critical. In this context, we test the role of particle's shape and material, using optically levitated nanodiamonds hosting NV centers to measure the particles' internal temperature and center-of-mass dynamics. We present a model to assess the nanodiamond internal temperature from its dynamics, adaptable to other particles.
 We also demonstrate that other mechanisms affect the nanodiamond dynamics and its stability in the trap.
Finally, our work, by showing levitating nanodiamonds as an excellent tool for studying nano-thermal effects, opens prospects for increasing the trapping stability of optically levitated particles. 
\end{abstract}

%%%%%%%%%%%%%%%%%%%%%%%%%%%%%%%%%%%%%%%%%%%%%%%%%%%%%%%%%%%%%%%%%%%%%
%                                                                   %
%                           FIGURE 1.                               %
%                                                                   %
%%%%%%%%%%%%%%%%%%%%%%%%%%%%%%%%%%%%%%%%%%%%%%%%%%%%%%%%%%%%%%%%%%%%%
\begin{figure*}[htb!]
    \centering
    \includegraphics[width=.99\textwidth]{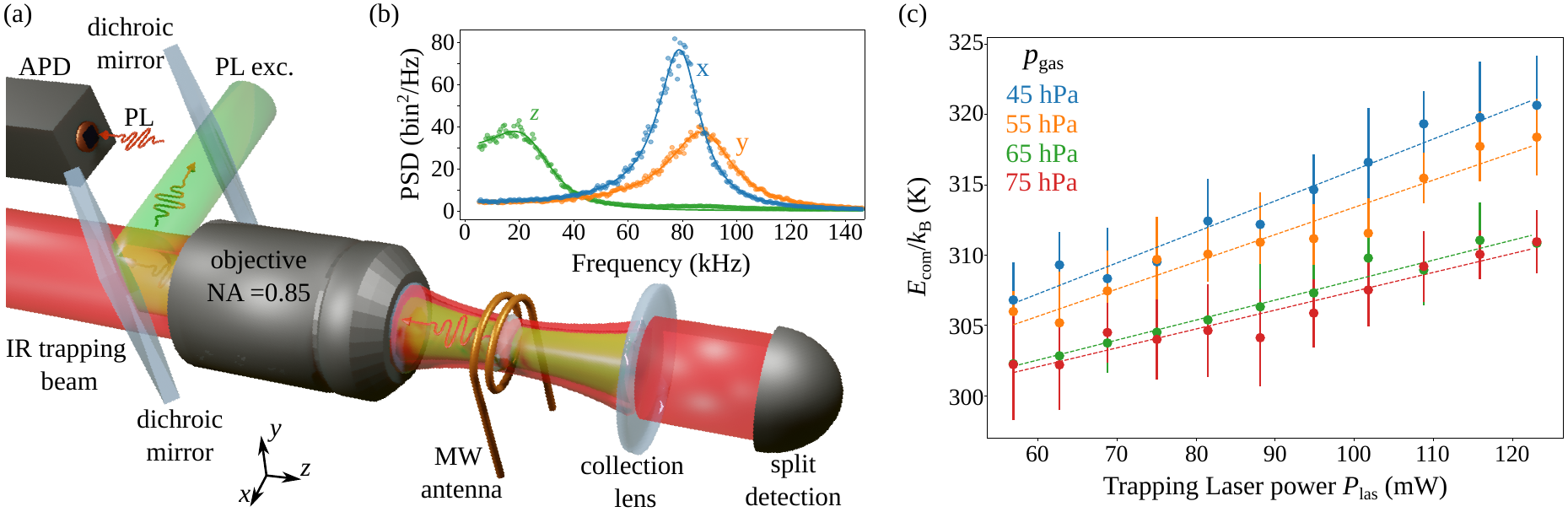}
    \caption{(a) Scheme of the experimental setup. A nanodiamond is trapped by a 1.55~$\mu$m laser (red) focused through a NA=0.85 objective, linearly polarized along the $y$-axis.  Its 3D motion is measured using a set of 3 split detectors (split detection). The NV centers hosted in the levitating diamond are excited by a $\lambda=532$~nm laser (green). The photoluminescence (PL) is collected by an avalanche photodiode (APD). The electron spin resonance spectrum of the NV centers is obtained thanks to a microwave field generated by a copper antenna placed around the levitating particle. (b) Typical power spectral densities (PSD) along the three axes $q={x,y,z}$ for a levitated nanodiamond at $p_\mathrm{gas}=45$~hPa. (c) Center-of-mass energy of a levitated nanodiamond as a function of the trapping laser power, for four different pressures. Error bars correspond to one standard deviation obtained from ten iterations of the experiment. Dotted lines are linear fits. }% used to determine the intercept required for the quantitative calibration of the energy.} 
    \label{fig:setup}
\end{figure*}

\section*{Introduction}
Providing efficient isolation from the environment, optical levitation in vacuum has proven to be an excellent system for investigations of quantum physics at the mesoscale~\cite{Delic2020S,Magrini2021N,Tebbenjohanns2021N}, for the tests of fundamental physical laws and weak forces sensing~\cite{Moore2020AHPPP},  and a unique testbed for stochastic thermodynamics~\cite{Gieseler2018E}. 
These exciting developments lie in the ability to control the particle's dynamics finely. Thus, understanding the parameters affecting this dynamics is of prime interest. Specifically, it is known that the particle's internal temperature may affect its dynamics, an effect known as \emph{hot Brownian motion}~\cite{Rings2010PRL,Falasco2014PRE} and recently studied for levitated silica nanospheres~\cite{millen2014nanoscale,Hebestreit2018PRA}. Nevertheless, this effect is expected to depend on the particles' shape and material%, and on the trapping medium
~\cite{Falasco2014PRE,Rodriguez-Sevilla2018AP}. 
Quantifying these effects is of fundamental interest for describing thermal effects at the nanoscale and for the development of nanothermometry techniques~\cite{millen2014nanoscale,frangeskou2018pure,Rahman2017NP}. 

To study the effect of nanodiamonds' internal temperature on their dynamics, we use optically levitated nanodiamonds doped with NV centers.
Nanodiamonds are particles with disparate shapes~\cite{Hoang2016PRL,rahman2016burning} that generally present important internal heating induced by laser absorption in levitation experiments~\cite{Hoang2016PRL,Delord2017APL}, an effect particularly significant for optical levitation due to the required high power trapping laser. NV centers are point defects of the diamond matrix, with unique spin and optical properties that can act as efficient temperature sensors~\cite{toyli2012measurement}. 
In the present paper, we first study the dynamics of optically trapped nanodiamonds in moderate vacuum (> 10~hPa) and assess the energy $E_\text{com}$ associated  with their center-of-mass motion. Then, we use NV centers thermometry to evaluate their internal temperature $T_\text{int}$.
By comparing these two quantities, we describe the coupling between nanodiamonds' internal temperature and dynamics and discuss how the diamond shape impacts it. These results allow us to assess the accommodation coefficient for nanodiamonds, a quantity challenging to measure at the single-particle level, underlining the power of levitodynamics for material science studies. Also, we introduce a simple method to quantify the internal temperature of a levitated nanodiamond from the measurement of its dynamics, with applications to nanothermometry of non-spherical nanoparticles~\cite{frangeskou2018pure,Rahman2017NP}.  Finally, we demonstrate that phenomena beyond particles' internal temperature impact the nanodiamond dynamics, potentially affecting the particle stability in the trap. Since mastering this particle stability is a critical task for developing diamond-based hybrid spin-mechanical levitated systems~\cite{Perdriat2021M}, our study also opens prospects beyond out of equilibrium physics of hot particles.
%%%%%%%%%%%%%%%%%%%%%%%%%%%%%%%%%%%%%%%%%%%%%%%%%%%%%%%%%%%%%%%%%%%%%
%                                                                   %
%                           FIGURE 2.                               %
%                                                                   %
%%%%%%%%%%%%%%%%%%%%%%%%%%%%%%%%%%%%%%%%%%%%%%%%%%%%%%%%%%%%%%%%%%%%%
\begin{figure*}[ht!]
\centering
\includegraphics[width=0.95\textwidth]{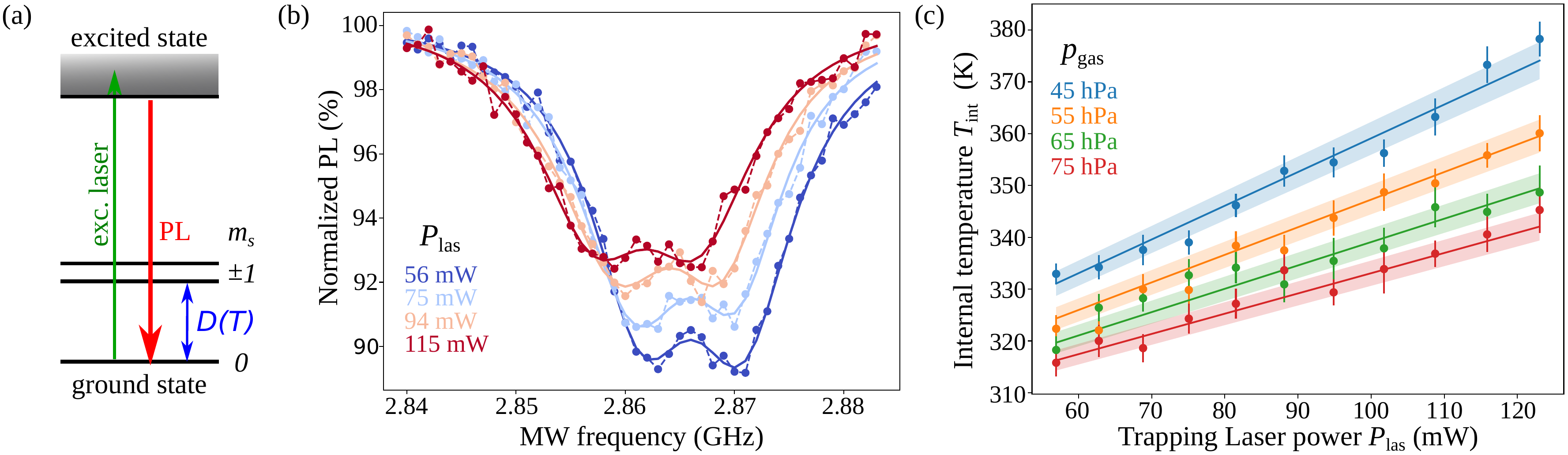}
\caption{(a) Scheme of a simplified NV center electronic structure. Under green laser excitation, NV centers emit bright photoluminescence centered around $\lambda_\text{PL}=700$~nm. The ground state of the NV center is a spin-triplet. The spin states $|m_s=0\rangle$ and $|m_s=\pm 1\rangle$ are split by a temperature dependant zero-field splitting $D(T)$. (b) Electron spin resonance spectra for the nanodiamond studied in figure~\ref{fig:setup}-(c), at a gas pressure $p_\mathrm{gas}=45$~hPa. Solid lines correspond to bi-Lorentzian fits of the spectra. Note that a small splitting between the state $|m_s=\pm 1\rangle$ is observed due to internal strain, a common feature in nanodiamonds. (c) Internal temperature of the levitated nanodiamond as a function of the infrared laser power, for four different pressures, extracted from ESRs in (b). Error bars correspond to the ESR fit uncertainties.  Solid lines correspond to a 2D fit using equation~(\ref{eq:Tint}), leading to 
$\kappa_\mathrm{heat} = 17.0 \pm 0.5$ K$\cdot$hPa$\cdot$mW$^{-1}$. Shadow areas indicate the uncertainties of the 2D fit.}
%Energy of the center-of-mass normalized to the room temperature fitting by a linear approximation. The computation of the slope gives access to the coefficient K which is linked to the coupling with the internal temperature. If the coupling is perfect and considering a spherical particle, this K coefficient need to be 0.33. b) the sum up of the computation of the K coefficient at each pressure. }
\label{fig:NV}
\end{figure*}

\section{Results and discussion}
To study the dynamics of optically levitated nanodiamonds, we use the experimental setup depicted in figure~\ref{fig:setup}-(a). A 100-nm nanodiamond (FND Biotech, brFND-100) is optically trapped inside a vacuum chamber, using a highly focused  $\lambda=1.55~\mu$m laser (Keopsys). The gas pressure in the chamber $p_\mathrm{gas}$ is finely controlled with a manual valve and measured with a capacitance gauge (Pfeiffer vacuum, CMR 361). 
We measure the nanodiamond dynamics using a standard common path interferometric detection scheme~\cite{Gieseler2012PRL}. The particle's forward scattered light is collected using a collimating lens and redirected onto three split detectors to measure the particle motion along the three axes $q=\{x,y,z\}$.  Typical power spectral densities (PSD) measured from a levitated nanodiamond are shown in figure~\ref{fig:setup}-(b). Assuming a harmonic motion in the underdamped regime, we fit these PSDs for each axis $q$ with the function~\cite{Hebestreit2017}:
\begin{equation}
        \mathcal S_{qq}(f)=\dfrac{A_q}{\pi} \frac{2 f_q^2\gamma_q}{(f^2-f_q^2)^2+f^2\gamma_q^2} \, ,
\end{equation}
and retrieve the natural frequency of the trap $f_q$, the damping rate $\gamma_q$, and the integrated power density $A_q$ of the motion. 
The energy of the center-of-mass motion $E_\mathrm{com}$ is proportional to the integrated power spectral density of the particle motion~\cite{Hebestreit2017}, such that, in the direction $q$,  $E_\mathrm{com}^q=\mathcal C_\mathrm{calib}^q A_q$.   $\mathcal C_\mathrm{calib}^q$ is a calibration factor related to the efficiency of the detection setup. It depends on our experimental configuration, detectors efficiency, and the laser measurement beam power. Here, we account for the latter by renormalizing the power spectral density by the laser power changes (see Supporting Information). 
Experimentally, the $z$-axis motion is at a lower frequency due to the weaker confinement along the optical axis. Thus, for most data sets, the $z$-frequency is smaller than the PSD linewidth, making its analysis challenging. Hence, we only address the particle motion along the $x$ and $y$ directions in the following.
In agreement with previous reports~\cite{Neukirch2015,hoang2016electron}, we observe that nanodiamonds are stably trapped down to a few tens of hPa and are generally lost from the trap at a pressure between 1 to 10~hPa.

Usually, one assumes that the particle is at room temperature to determine the calibration factor $\mathcal C_\mathrm{calib}^q$ and quantitatively determine the center-of-mass energy. Under this assumption, the center-of-mass energy is the thermal energy $E_0=k_B T_0$, where $k_B$ is the Boltzmann constant and $T_0=294$~K is the room temperature. In our experiment, since the center-of-mass energy is expected to increase with the laser power, the particle is inherently out-of-equilibrium. While alternative calibration methods have been used~\cite{Millen2020RPP}, we propose here a novel and simple calibration approach to overcome the difficulties imposed by the non-equilibrium situation.  We thus record the dynamics of the levitated nanodiamond for increasing laser powers. 
We then compute and fit the associated PSDs to determine the value of $A_q$ as a function of the trapping laser power (see Supporting Figure~S1).
To reduce uncertainties, we average the values of $A_q$ over ten successive measurements while ramping the laser power. Finally, from a linear extrapolation to vanishing laser powers, where we expect the particle to be at equilibrium with its environment, we determine the calibration factor:
\begin{equation}
        \mathcal C_\mathrm{calib}^q =  \frac{E_0}{A_q(P_\mathrm{las}=0)}\, .
\end{equation}
Note that a benefit from this procedure is that we independently determine the calibration factor at each pressure, overcoming calibration uncertainties related to experimental drifts induced by pressure changes inside the chamber~\cite{Hebestreit2018RoSI}.

From this calibration, we can determine the center-of-mass energy of the particle. We observe a rise of the center-of-mass energy with increasing laser power and decreasing gas pressures, as plotted in figure~\ref{fig:setup}-(c). This effect can be attributed to an increase of the nanodiamond internal temperature, which will affect its dynamics, as expected for hot Brownian motion~\cite{millen2014nanoscale}. 
To verify this hypothesis and quantify the link between the observed effect  and the nanodiamond internal temperature, we measure this temperature using NV color centers hosted in the diamond matrix~\cite{hoang2016electron,Pettit2017JOSABJ,Delord2017APL}.
%
%\section{Internal temperature of levitated nanodiamonds}
The trapped nanodiamonds are heavily doped with thousands of NV color centers. 
Its electronic structure, displayed in figure~\ref{fig:NV}-(a), provides a bright and stable red photoluminescence (PL) under green laser excitation. In addition, its ground state is a spin-triplet $S=1$, where the exact splitting $D(T)$ between the spin sublevels $|m_s=0\rangle$ and $|m_s=\pm 1\rangle$ is temperature-dependent, its dependence being empirically described by a polynomial function~\cite{toyli2012measurement,hoang2016electron,Delord2017APL}. Thus, NV centers can be used as sensitive thermometers. The PL being spin-dependent, we can optically detect the electron spin resonance (ESR) of the NV centers and precisely determine the evolution of $D(T)$ with temperature. 
Experimentally, we excite the NV centers in the levitated nanodiamond using  a green laser (Laser Quantum GEM, $\lambda=532$~nm, laser power $P_\mathrm{green}\lesssim 100$~µW$\ll P_\mathrm{las}\approx 100$~mW) through the trapping objective. The weak excitation power that we use prevents heating of the nanodiamond by the green laser absorption. The related photoluminescence (PL) is collected through the same objective, split from the trapping and the green excitation lasers using two dichroic mirrors, and sent onto an avalanche photodiode (APD) in the single-photon counting regime. A copper wire rolled around the optical trap acts as a microwave antenna (fig.~\ref{fig:setup}-(a)). 
We record ESR spectra by measuring the PL of the NV centers while sweeping the applied microwave frequency for different trapping laser power. As shown in figure~\ref{fig:NV}-(b), we observe a clear shift of the ESR spectra, highlighting the nanodiamond crystal's heating. From the empirical law proposed by Toyli~\textit{et al.}~\cite{toyli2012measurement}, we quantitatively assess the internal temperature of the levitated nanodiamond. 
Given the small size of the nanodiamond and its high thermal conductivity, we assume that its internal temperature is homogeneous and corresponds to the particle surface temperature.  
In the pressure regime explored here, the internal temperature arises from the competition between  absorption of the trapping laser and thermal conduction~\cite{hoang2016electron}, such  that it scales as
\begin{equation}
        T_\mathrm{int} = T_0+ \kappa_\mathrm{heat} \frac{P_\mathrm{las}}{p_\mathrm{gas}}\, , 
    \label{eq:Tint}
\end{equation}
as demonstrated in figure~\ref{fig:NV}-(c).
A fit acting on both gas pressure and laser power, allows us to extract the heating rate $\kappa_\mathrm{heat}$ and $T_0$. 
In practice, internal strain in the nanodiamonds may shift the value of the zero-field splitting $D(T)$, leading to a definition of the internal temperature $T_\text{int}$ within a constant. This effect is corrected by shifting the measured temperatures such that the value of $T_0$ returned by the fit is equal to the room temperature~\cite{hoang2016electron}. Data and fits in figure~\ref{fig:NV}-(c) are corrected from this effect and present quantitative values of the diamond internal temperature.

%%%%%%%%%%%%%%%%%%%%%%%%%%%%%%%%%%%%%%%%%%%%%%%%%%%%%%%%%%%%%%%%%%%%%
%                                                                   %
%                           FIGURE 3.                               %
%                                                                   %
%%%%%%%%%%%%%%%%%%%%%%%%%%%%%%%%%%%%%%%%%%%%%%%%%%%%%%%%%%%%%%%%%%%%%
\begin{figure}[htb!]
        \centering
        \includegraphics[width=0.95\columnwidth]{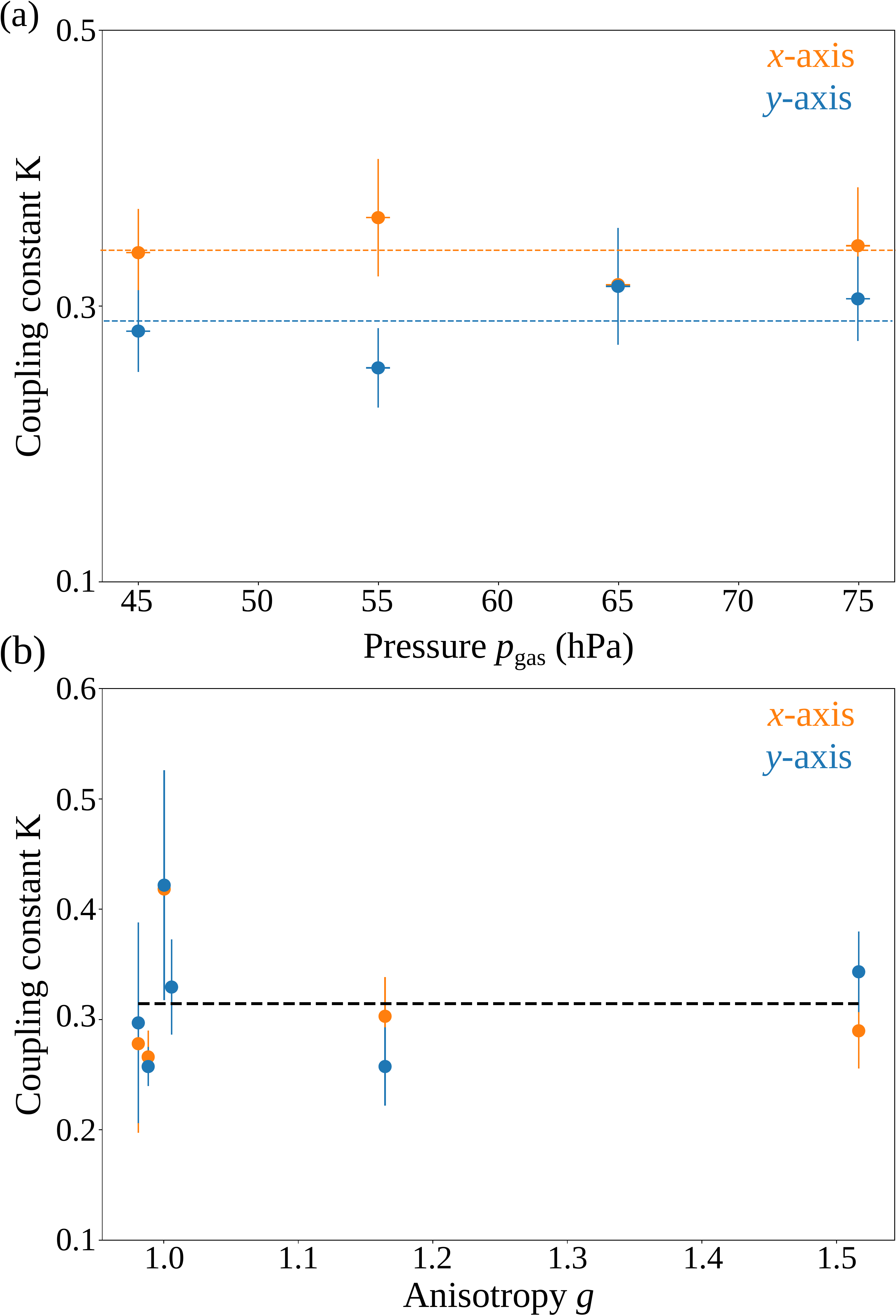}
        \caption{(a) Hot Brownian motion coupling constant $K$ along the axes $x$ (blue) and $y$ (orange) for the nanodiamond presented in figures~\ref{fig:setup} and~\ref{fig:NV}. Dotted lines indicate the averaged values of $K_x=0.34\pm0.04$ and $K_y=0.28\pm0.03$. Error bars on the pressure correspond to typical pressure drifts during data acquisition. (b) Hot Brownian motion coupling constant $K$ along the axes $x$ (blue) and $y$ (orange), as a function of the anisotropy factor $g=\gamma_x/\gamma_y$, for a set of nanodiamonds. The black dotted line corresponds to the averaged $K$ over all particles and axes.}

\label{fig:K}
\end{figure}

%\section{Internal temperature impacts on the nanodiamond dynamics}
Thus, our experimental setup allows us to simultaneously assess the energy of the nanodiamond center-of-mass and its internal temperature. To quantify the coupling between these two quantities, we introduce the coefficient $K$ such that
\begin{equation}
K = \dfrac{\Delta E_\mathrm{com}}{k_B \Delta T_\mathrm{int}}\, , % = K A_\mathrm{heat}\frac{P_\mathrm{las}}{p_\mathrm{gas}}
\end{equation}
with $\Delta E_\mathrm{com}$ the increase in center-of-mass energy corresponding to an increase in internal temperature $\Delta T_\text{int}$.
Since we have shown that, under our experimental conditions, both the center-of-mass energy and internal temperature increase linearly with laser power, $K$ is independent of the laser power. Also, it corresponds to the ratio of slopes of the curves in figures~\ref{fig:setup}-(c) and~\ref{fig:NV}-(c). As an example, figure~\ref{fig:K}-(a) presents the value of this coupling $K$ for the nanodiamond used in figures~\ref{fig:setup}-(c) and~\ref{fig:NV}-(c). 

It is interesting to note that the value of $K$ is independent of the gas pressure and relatively similar for the two axes $x$ and $y$. 
This result can be discussed in light of the two baths formalism introduced by Millen \textit{et al.}~\cite{millen2014nanoscale} to describe the coupling of the internal temperature of a spherical levitated particle to its center-of-mass motion. In this model, the coupling is mediated by the interaction of the surrounding gas on the hot particle surface. Initially with thermal energy associated with the room temperature $T_0$,  the impinging gas molecules on the hot particle will see the accommodation of their thermal energy when scattered by the hot particle surface. In return, these emerging heated particles will contribute to the particle Brownian motion, as an ancillary thermal bath of effective emerging temperature $T_\text{em}>T_0$. The strength of this effect is quantified through the thermal accommodation coefficient $\alpha_c=\frac{T_\text{em}-T_0}{T_\text{int}-T_0}$, which describes the average fraction of the surface temperature which is transferred to the gas molecules during a collision with the hot particle. 
%proportion of gas molecule that accomodate their temperature with the hot particle during a collision.
It varies from $\alpha_c=0$, where all gas molecules are elastically reflected, to $\alpha_c=1$, where all gas molecules get an increase of their thermal energy to $k_B T_\text{int}$ after the collision. As such, it is a convenient coefficient to describe the interaction of the particle with its environment.
Nevertheless, it depends on the particle materials and the surrounding gas, and its determination at the single nanoparticles level is generally challenging. 
In the case of levitated silica spheres in vacuum, Hebestreit~\textit{et al.} presented an approach to estimate the accommodation coefficient of nano-silica, inferring the internal temperature of the levitated particle from the thermal changes of its physical properties and its relaxation dynamics after a controlled heating~\cite{Hebestreit2018PRA}. However, this characterization can not be straightforwardly applied to any particles.
In our experiment, knowing the coupling constant $K$, we can estimate the accommodation coefficient $\alpha_c$ from a linearisation of the two bath model, leading to $K =\frac{\pi}{\pi+8} \alpha_c \approx 0.28 \alpha_c$ (see Supporting Information). Thus, from the fits in figure~\ref{fig:K}-(a), we find a nanodiamond accommodation coefficient $\alpha_c^x = 1.01 \pm 0.12$ for the $x$-axis and $\alpha_c^y = 1.21 \pm 0.13$ for the $y$-axis. Interestingly, we find values very close to the maximal value of $\alpha_c=1$, corresponding to the largest coupling possible between the particle Brownian motion and its internal temperature.
While we could expect such an excellent thermal accommodation coefficient for bulk diamond, the quality of the diamond surface, \textit{i.e.}, graphitization~\cite{rahman2016burning}, may drastically reduce the accommodation, an effect we do not observe.
Also, we note that accounting for the uncertainties, we observe thermal accommodation coefficients that are slightly above 1 and not isotropic ($\alpha_c^x\neq\alpha_c^y$). We explain this effect because the model is designed for spherical particles. At the same time, the used nanodiamonds, made from bulk diamond milling, are known to present very irregular shapes~\cite{Hoang2016PRL,rahman2016burning}, a parameter that is expected to impact the coupling of the levitated particle to the surrounding gas molecules~\cite{Martinetz2018PRE} (see Supporting Information).

To study further the impact of the particle shape on the coupling coefficient $K$, we reproduce the previously described procedure on a batch of nanodiamonds. To characterize the shape of these nanodiamonds, we determine, from the PSD fits, the damping rate $\gamma$ of the particle along the $x$ and $y$ axes. This damping arises from a Stokes drag and is related to the particle effective surface in the particle motion direction~\cite{Martinetz2018PRE}. 
While not providing quantitative information about the exact particle shape, the measurement of the ratio $g=\gamma_x/\gamma_y$ is a good indicator of the particle anisotropy. As a good example, the libration mode of levitated nanodiamonds has only been observed for particles with $g$-ratio away from one~\cite{Hoang2016PRL}. Also, note that we only observe $g\geq 1$ due to the orientation of the longer axis of the particle along the polarization direction~\cite{Hoang2016PRL}.

We plot in figure~\ref{fig:K}-(b) the coupling constant $K$ as a function of the anisotropy $g$-factor. We observe slight variations on the measured values of $K$ and the discrepancy between axes remains smaller than the dispersion among different particles. Accounting for this dispersion, both over the studied particles and axes, we estimate an average value of $\bar K=0.31\pm0.04$. 
%corresponding to $\bar \alpha_c=1.1\pm0.2$.
Therefore, for most investigated nanodiamonds, the coupling between internal temperature and center-of-mass dynamics can be relatively well described by the spherical particle formalism with an accommodation coefficient close from $\alpha_c^\mathrm{ND}=1$. Also, the relatively small dispersion we observe for the value of $K$ may be used to provide a reasonable estimate of any nanodiamond internal temperature from its dynamics using a Hot Brownian coupling constant $\bar K=0.31$. This could be particularly helpful to assess the absorption of particles that are not hosting spin-active defects~\cite{frangeskou2018pure}. 

%%%%%%%%%%%%%%%%%%%%%%%%%%%%%%%%%%%%%%%%%%%%%%%%%%%%%%%%%%%%%%%%%%%%%
%                                                                   %
%                           FIGURE 4.                               %
%                                                                   %
%%%%%%%%%%%%%%%%%%%%%%%%%%%%%%%%%%%%%%%%%%%%%%%%%%%%%%%%%%%%%%%%%%%%%
\begin{figure}[htb!]
\centering
\includegraphics[width=0.95\columnwidth]{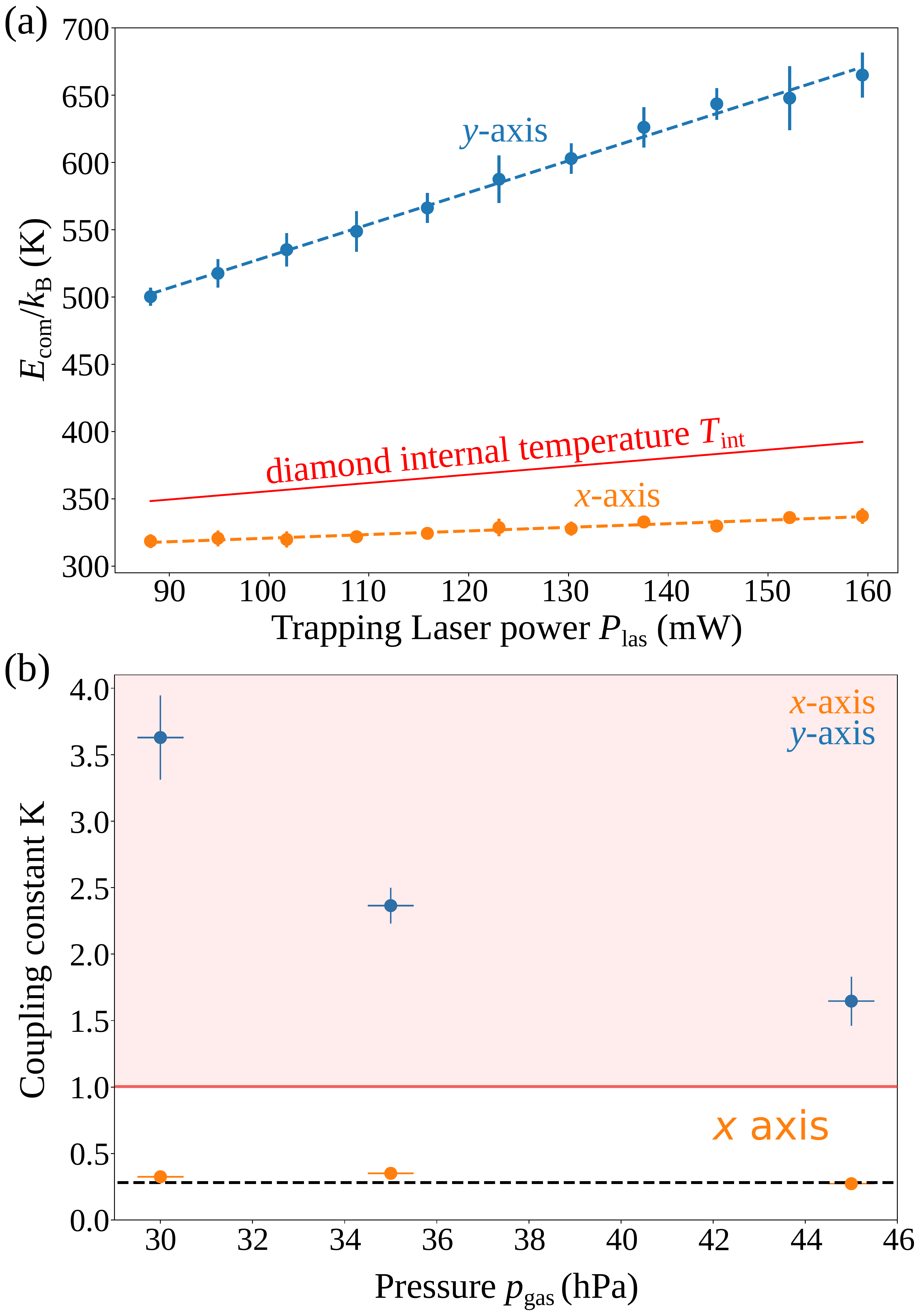}
\caption{(a) Center-of-mass energy as a function of the laser power for a particle with an unexpected behavior along the $y$-axis. (b) Associated hot Brownian motion coupling constant $K$  along the $x$-axis (orange) and $y$-axis (blue). The black dotted line corresponds to the average $K$ for the figure~\ref{fig:K}-(b) nanodiamonds. The red shadow area corresponds to the case $K>1$ that can not be explained in the model used in this work. Error bars on $K^x$ are smaller than the size of the symbols.}
\label{fig:overH}
\end{figure}

While this proposed model applies to six over the eight studied particles, we observe an unexpected anisotropic over-heating of the diamond center-of-mass energy for the remaining two.
%We have observed this behaviour at least for two particles over the height investigated.
Figure~\ref{fig:overH}-(a) shows the measured energy of the center-of-mass for one of these particles. While estimating the coupling coefficient $K$, it appears that it reaches values far above one, along the $y$-axis, while keeping an expected value of $K^x\approx0.3$ along the $x$-axis. Also, as shown in figure~\ref{fig:overH}-(b), the value of $K$ along the $y$-axis increases while the gas pressure is reduced, conversely to what we observe for the $x$-axis or the particles in figure~\ref{fig:K}.
The measured  $K^y>1$ is definitive proof that the increase in center-of-mass energy can not be explained only by the diamond internal temperature  and is proof that other heating mechanisms exist.
The fact that we reproduce each measurement ten times at each pressure rules out a time-dependent change, such as a particle degradation~\cite{rahman2016burning} that would modify our measurement over time. In addition, these "\emph{over-heated}" particles present neither a size, a laser absorption ($\kappa_\mathrm{heat}$), nor an anisotropy ratio $g$ further away from the particles shown in figure~\ref{fig:K}-(b) (see Supporting Information).
Recently, center-of-mass heating mechanisms have been identified for levitated silica~\cite{Svak2018NC} and birefringent~\cite{Arita2020SA} micro-particles involving non-conservative interactions. However, in both cases, the particle is coherently driven, leading to a reduction of the PSDs linewidths, an effect we do not observe here.
Alternatively, the random shape of the nanodiamonds may also induce non-trivial dynamics. Recent numerical studies have highlighted that optical trapping may become unstable for given random particle shapes, even for particle sizes small compared to the laser wavelength~\cite{Herranen2019PO}.
A complete understanding of this effect would require determining the exact particle shape, either by on-demand trapping or using nanofabricated diamond particles of controlled shapes and aspect ratios~\cite{Radtke2019M}. Besides, it would be beneficial to monitor and control most of the degrees of freedom of the particle dynamics~\cite{Bang2020PRR} since the complex dynamics potentially arises from the coupling of different degrees of freedom.
Finally, beyond the nature of this heating mechanism, which is not driven by the particle's internal temperature, an important question is to know if this phenomenon exists for all nanodiamonds but with variable strengths. In such a case, reducing the absorption of the nanodiamond, e.g., using ultrapure diamond~\cite{frangeskou2018pure}, will not be sufficient to keep levitated nanodiamonds dynamics at equilibrium with its environment, thus eventually limiting its stability in the trap.  

\section*{Conclusion}
Taking advantage of the unique spin features of NV color centers in optically levitated nanodiamonds, we simultaneously measure these nanodiamonds' internal temperature and center-of-mass energy.
We then show that the coupling between these quantities is only weakly affected by the shapes of nanodiamonds and corresponds to a hot Brownian coupling constant $\bar K=0.31\pm0.04$ in our model. Equivalently, it can be well described by the two baths formalism introduced by Millen \textit{et al.}~\cite{millen2014nanoscale} for spherical particles,  with an accommodation coefficient for the nanodiamond surface close to $\alpha_c^\mathrm{ND}=1$.
These results provide a simple way to assess the internal temperature of non-spherical nanodiamonds from the measurement of their Brownian dynamics with moderate errors. % and can potentially be extended to other materials.
They also demonstrate NV doped nanodiamonds as a powerful tool to study thermal effects at the nanoscale. A natural extension of our work would be investigating the accommodation coefficient dependence on the particle size~\cite{Wang2020PRE} or the hot Brownian motion for other degrees of freedom such as rotation and libration~\cite{Rodriguez-Sevilla2018AP,Martinetz2018PRE}.

Besides, we also report a new center-of-mass heating mechanism that is anisotropic and not driven by the internal temperature. We observe this mechanism on only a fourth of the studied particles, and we attribute it to a complex dynamics linked to the nanodiamonds' random shapes. 
We highlight that this heating mechanism may contribute to the nanodiamond exit from the optical trap at medium vacuum pressure, along with the particles' internal heating. Thus, careful understanding and control of the internal temperature and of the center-of-mass and rotational motion of the levitated nanodiamonds are required to reach higher vacuum conditions and unleash the potential of levitated nanodiamonds for quantum spin-mechanics experiments~\cite{Perdriat2021M}.

\section*{Acknowledgments}
We thank E.~Fayen and C.~Fournier for their seminal contribution to experimental developments.
\section*{Funding Sources}
This work is supported by the Investissements d’Avenir of LabEx PALM (ANR-10-LABX-0039-PALM), by the Paris \^ile-de-France Region in the framework of DIM SIRTEQ, and by the ANR OPLA project (ANR-20-CE30-0014).
 
%\begin{suppinfo}
%This material is available free of charge via the internet at http://pubs.acs.org.
%Advanced discussions on center-of-mass energy calibration, modeling of hot Brownian motion, and details on the studied particles (PDF). 
%\end{suppinfo}

\bibliography{biblio}

\appendix

\newpage
\onecolumn

\section{Supporting Information}

\makeatletter
\renewcommand{\thefigure}{S\arabic{figure}}
\setcounter{figure}{0}
\renewcommand{\theequation}{S\arabic{equation}}
\setcounter{equation}{0}
\renewcommand{\thepage}{S\arabic{page}}
\setcounter{page}{1}
\renewcommand{\thetable}{S\arabic{table}}
\makeatother

\renewcommand{\thefootnote}{$\star$}

\subsection{Calibration of the particle displacement}%
\label{sec:time_trace_analysis}

The standard procedure for determining a levitated particle center-of-mass energy is well described by Heberstreit et al.~\cite{Hebestreit2018RoSI}. 

\noindent In brief, we measure the particle dynamics along the $q$-axis through an optical signal $V_q=c_\text{calib} q$, where $c_\text{calib}$ is a setup-dependent calibration factor. 

Assuming that the optical trap is harmonic\footnote{Due to trap anharmonicities, our procedure underestimates the center-of-mass energy by a few percent~\cite{Hebestreit2018RoSI}. We neglect this effect, since the error is lower than our measurement uncertainties.}, and by analogy with the equipartition theorem, we defined the center-of-mass energy of the particle $E^q_\text{com}=k_B T_\text{com}$, where  $T_\text{com}$ is the effective temperature of the particle center-of-mass motion, such that 
\begin{equation}
        \dfrac{1}{2}k_B T_\text{com} =  \dfrac{1}{2} m \Omega^2 \langle q^2 \rangle = \dfrac{1}{2} m \Omega^2 \dfrac{ \langle V_q^2 \rangle }{c_\text{calib}^2} 
\end{equation}
with $m$ the particle mass, and $\Omega$ the angular frequency of the harmonic trap. 

Furthermore, as shown in section 4.1 of reference~\cite{Hebestreit2017}, under the harmonic oscillator  approximation, the power spectral density (PSD) $S_{qq}$ of the signal $V_q$ associated with the particle position writes:
\begin{equation}
        \mathcal S_{qq}(f)=\dfrac{A_q}{\pi} \frac{2 f_q^2\gamma_q}{(f^2-f_q^2)^2+f^2\gamma_q^2} \, ,
        \label{eq:PSD}
\end{equation}
where $f_q=\frac{\Omega_q}{2\pi}$ is the trap frequency and $\gamma=\frac{\Gamma}{2\pi}$ is the reduced damping. Then, one can show that $A_q = \langle V_q^2 \rangle=c_\text{calib}^2 \langle q^2\rangle$, such that
\begin{equation}
 E^q_\text{com}=m\Omega_q^2 \dfrac{A_q}{c_\text{calib}^2}\, .
 \label{eq:Ecom1}
\end{equation}
As discussed in the main text, the values of $A_q$ and $\Omega_q=2\pi f_q$ can be obtained from a fit of the experimental PSD by the equation~(\ref{eq:PSD}).

To determine the center-of-mass energy, we need to assess the value of $c_\text{calib}$.  In the literature, this is generally done from a reference experiment.
However, in the present case, we have to account for the calibration factor dependence on the measurement laser power and changes in the detection sensitivity that may arise with setup drifts associated with gas pressure changes~\cite{Hebestreit2018RoSI}.
In the following sections, we address these questions.

\subsubsection{Impact of the laser power on calibration}%
\label{sub:correcting_for_laser_power_change}
Our measurement scheme of the particle motion uses a common path interferometer. We use the same infrared laser for the measurements than for trapping the particle. We note $P_\text{las}$ the power of this laser. Thus, the displacement signal $V_q$ corresponds to the interference between the field scattered by the particle $E_s$ with the reference field directly transmitted through the chamber $E_\text{ref}$. Since both $E_s$ and $E_\text{ref}$ scale as $\sqrt{P_\text{las}}$, the measurement sensitivity $c_\text{calib}$ is then proportional to $P_\text{las}$.

We note $c_\text{calib} =c_0 P_\text{las}$, where $c_0$ is independent of the laser power, and equation~(\ref{eq:Ecom1}) becomes:
\begin{equation}
        E^q_\text{com} = \dfrac{m\Omega_q^2}{c_0^2} \dfrac{A_q}{P_\text{las}^2} \, .
        \label{eq:Ecom2}
\end{equation}

Also, the trap angular frequency scales as the square root of the trapping laser power, $\Omega_q = \beta_q \sqrt{P_\text{las}}$, with $\beta_q$ a constant defined by the physical properties of the trap~\cite{Gieseler2012PRL}. 

Thus, we simplified the expression of the center-of-mass energy as~: 
\begin{eqnarray}
        E^q_\text{com} &=& \dfrac{m}{c_0^2} \dfrac{A_q}{\Omega_q^2} \\
                       &=&  \dfrac{m\beta^4}{c_0^2} \dfrac{A_q}{4\pi^2 f_q^2} \\
                       &=&  \dfrac{ \mathcal C_\text{calib}}{f_q^2}\cdot A_q\, ,
\end{eqnarray}
where $\mathcal C_\text{calib}= \dfrac{m\beta^4}{4\pi^2  c_0^2}$ is the detection calibration factor normalized to account for the laser power dependence on the detection sensitivity. Thus, it only accounts for drifts of the detection sensitivity when changing the gas pressure. The dependence of the detection sensitivity on the laser power corresponds to the term $1/f_q^2$. To easily account for it, we introduce the integrated power spectral density normalized by the change in laser power $\tilde A_q = \dfrac{A_q}{f_q^2}$, which is the quantity actually used in the main text. 

\subsubsection{Quantitative assessment of the calibration factor}%
\begin{figure}[htb]
        \centering
        \includegraphics[width=0.8\linewidth]{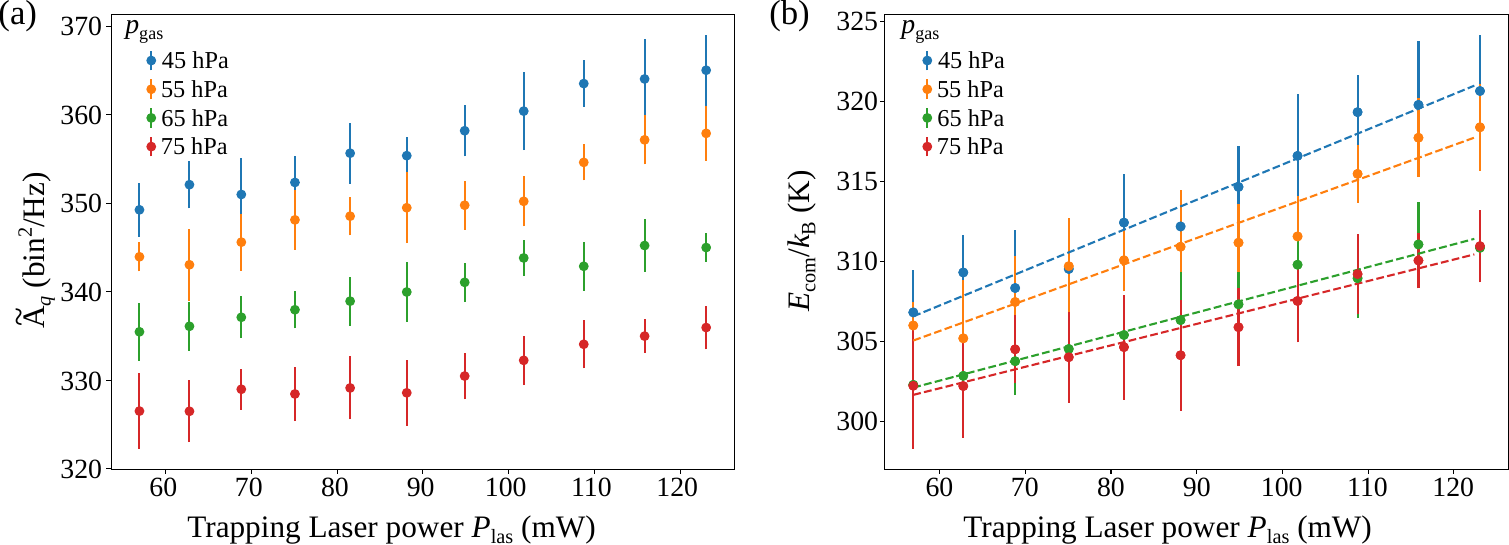}
        \caption{Quantitative assessment of the calibration factor. (a) Direct measurement of $\tilde A_q$. Error bars correspond to the standard deviation measured over ten measurements. (b) $E_\text{com}$ computed after calibration of the data in (a). Dotted lines correspond to linear fits.}%
        \label{fig:Calib}
\end{figure}

As quickly discussed in the main text, we use the following procedure to determine the particle center-of-mass energy:
\begin{enumerate}
        \item First, we determine the values of $\tilde A_q= \dfrac{A_q}{f_q^2}$ from the fit of PSD of the particle trajectories. Then, we measure ten times $\tilde A_q$ as a function of the laser power, at each pressure. 
        \item We fit  $\tilde A_q$ as a function of $P_\text{las}$ by a linear function.  From the $y$-intercept of the fit, we extrapolate $\tilde A_q(0)$.
        \item Assuming that the particle is at thermal equilibrium with its environment for vanishing laser power, we compute the calibration factor $\mathcal C_\text{calib}=\dfrac{k_B T_0}{\tilde A_q(0)}$. $\mathcal C_\text{calib}$ is determined independently at each pressure, avoiding biases induced by setup drifts with pressure changes. 
        \item Finally, we plot $E_\text{com}^q=\mathcal  C_\text{calib}\cdot \tilde A_q$, as shown in figure~\ref{fig:Calib}-(b).

\end{enumerate}

\subsection{Linearisation of the two bath model}%
\label{sec:linearisation_of_the_two_bath_model}

Following Millen et al.~\cite{millen2014nanoscale}, the center-of-mass temperature of a spherical particle writes:
\begin{equation}
        T_\text{com} = \dfrac{T_0^{3/2}+\frac{\pi}{8}\left(T_0+\alpha_c \Delta T_\text{int}\right)^{3/2}}{T_0^{1/2}+\frac{\pi}{8}\left(T_0+\alpha_c \Delta T_\text{int}\right)^{1/2}}\, , 
\end{equation}
where $T_0$ is the ambient temperature, $\Delta T_\text{int}=T_\text{int}-T_0$ is the internal temperature increase, and $\alpha_c$ the accommodation coefficient. One can rewrite it:
\begin{eqnarray}
        T_\text{CoM} &=& T_0 \dfrac{1+ \frac{\pi}{8}\left(1+\alpha_c \frac{\Delta T_\text{int}}{T_0}\right)^{3/2}}{1+ \frac{\pi}{8}\left(1+\alpha_c \frac{\Delta T_\text{int}}{T_0}\right)^{1/2}} \\
                     &=& T_0 \left(1 + \dfrac{\pi}{\pi+8}\alpha_c \dfrac{\Delta T_\text{int}}{T_0}+ \dfrac{4\pi}{(\pi+8)^2}\alpha_c^2\left(\dfrac{\Delta T_\text{int}}{T_0}\right)^2+o \left(\dfrac{\Delta T_\text{int}}{T_0}\right)^2\right)
\end{eqnarray}
for moderate increases of internal temperature such that $\dfrac{\Delta T_\text{int}}{T_0}\ll1$. 

Experimentally, fitting the data with the whole equation or the simple linear approximation returns the same value of the accommodation coefficient within the uncertainties. Thus, we assume that the linear approximation is verified in the main text. In such, a case we have
\begin{equation}
        \Delta T_\text{com} = \dfrac{\Delta E_\text{CoM}}{k_B} \approx \alpha_c \dfrac{\pi}{\pi+8}\Delta T_\text{int} 
\end{equation}
and we note $K= \alpha_c \dfrac{\pi}{\pi+8}$.

\subsection{Particle shape impacts on the hot Brownian constant $K$}%
To illustrate how the shape of the particle affects its hot Brownian motion, we propose to discuss the simple case of cylinder-like particles. 

The effect of surface temperature for cylinders has been investigated by Martinetz et al.~\cite{Martinetz2018PRE}. We note that the definition of the accommodation coefficient $\alpha_c$ by Martinetz et al. slightly differs from the one used in the main text of this work. Nevertheless, the two definitions are equivalent for $\alpha_c=1$, which we demonstrate to be the case for nanodiamonds. Thus we assume  $\alpha_c=1$ in the following.

\begin{figure}[htpb]
        \centering
        \includegraphics[width=\linewidth]{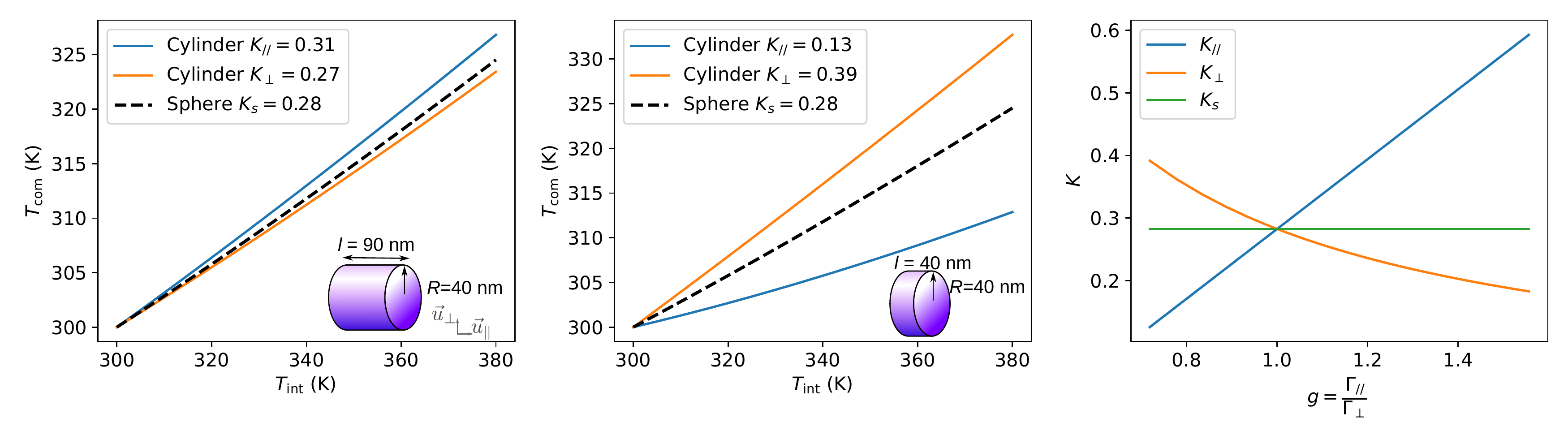}
        \caption{Effect of the cylinder shape on the hot Brownian motion. (a) and (b) Center-of-mass temperature $T_\text{com}$ of a cylinder of internal temperature $T_\text{int}$, radius $R = 40$~nm, and length $l=90$~nm (a) or $l=40$~nm (b).  The center-of-mass temperature of the motion along the cylinder axis (blue) or orthogonal to it (orange) are compared to the case of a sphere (black dotted line).  (c) Hot Brownian motion coefficient $K$ as a function of the cylinder damping anisotropy $g$. The values of $K$ are obtained from the slope in (a) and (b). We note that $g=1$ corresponds to $l=2R$.}%
        \label{fig:KsimuSI}
\end{figure}

Following the work by Millen et al.\cite{millen2014nanoscale}, we split the gas contribution in two independent heat baths of different temperatures. We thus compute the effective center-of-mass temperature $T_\text{com}$ for a hot particle of surface temperature $T_\text{int}$. 
Figures~\ref{fig:KsimuSI}-(a) and (b) show the resulting center-of-mass temperatures for cylinders of radius $R=40$~nm and respective lengths $l=90$ and $l=40$~nm. 
From the slope of these curves, we estimate the value of the hot Brownian motion constant $K=\dfrac{\Delta T_\text{com}}{\Delta T_\text{int}}$. As an example, figure~\ref{fig:KsimuSI}-(c) shows the values of $K$ for a motion along the two symmetry axes of the cylinders as a function of the anisotropy factor $g=\dfrac{\Gamma_{\perp}}{{\Gamma_{//}}}$, which is varied by changing $l$ while keeping $R=40$~nm. 

From this simple model, we highlight that:
\begin{itemize}
        \item Depending on the cylinder geometry, the value of $K$ may strongly vary and is axis dependant. 
        \item For isotropic damping (\textit{i.e.}, an anisotropy factor $g=\Gamma_x/\Gamma_y\approx1$), the hot Brownian motion is equivalently described by the one of a sphere. This is true at least for other simple geometries such as the cuboid.
        \item For a non-isotropic particle, we generally observe a value of $K$ along one of the cylinder symmetry-axes larger than the expected value for a sphere and smaller along the other cylinder symmetry axis. 
\end{itemize}
These results match well our experimental observations. We always measure $K_x$ and $K_y$ on a different side of the expected $K_s\approx0.28$ for a sphere. The dispersion in $K$ is moderate, which is in good agreement with the fact that most particles have a weak damping anisotropy ($g\approx1$). This confirms that for a diamond sphere the expected value of $K$ will be $K_s\approx0.28$, which leads to an accommodation coefficient for nanodiamonds of $\alpha_c\approx1$. 

Also, we note that the particle asymmetry provides a better coupling to the heat bath than for a sphere ($K_x>K_s$). So, assuming that the particle is a sphere leads to an effective accommodation coefficient larger than one, as discussed in the main text, even if $\alpha_c \geq 1$ is not physically meaningful.

Finally, in contrast with the cylinder case, we do not observe a significant deviation from the sphere model in our experiments, even for the larger measured values of $g\approx1.5$. This highlights that the cylinder model is not well suited for our nanodiamonds that are of more irregular shapes. To test the model of hot Brownian motion of a cylinder, it would be very interesting to trap diamond nanopillars of well control aspect ratio.

\subsection{Characterization of overheated particles}%
We study height particles, from which two present an unexpected behavior in terms of heating of their center-of-mass motion. 
Here, we compare some physical parameters measured for these two particles to the value obtained by averaging over all the other particles. 
The given errors correspond to the fit uncertainties for the \textit{overheated particles}, and to the standard deviation in the averaged values. 

We compare
\begin{itemize}
        \item 
the hydrodynamic radius, which is the radius that would have a diamond sphere of the same damping $\Gamma_x$ as the particle~\cite{hoang2016electron}, \textit{i.e.}
\begin{equation}
r_x^\text{hydro} = 0.619 \dfrac{9}{\sqrt{2\pi}\rho_\text{diam}}\sqrt{\dfrac{M}{N_A k_B T_0}}\dfrac{p_\text{gas}}{\Gamma_x}\, ,
\end{equation}
where $\rho_\text{diam}\approx 3500$~Kg/m$^3$ is the diamond density, $M$ the molar mass of air, $T_0$ the room temperature and $p_\text{gas}$ the pressure inside the vacuum chamber.
While not giving a precise physical quantity, the hydrodynamic radius provides an idea of the typical particle size.
\item the anisotropy ratio $g=\Gamma_x/\Gamma_y$.
\item the heating rate $\kappa_\text{heat}$.
\end{itemize}
For all these quantities, the \textit{overheated} particles are in line with the expectation of the whole batch of particles, as shown in Table~\ref{tab:Carac}.

\begin{table*}[htpb]
        \centering
        \begin{tabular}{ |p{3.5cm}| p{3.5 cm} | p{3.5cm}| p{3.5cm} | } 
  \hline
  & overheated part. 1 & overheated part. 2 & all particles \\ 
  \hline
        \footnotesize{$r^\text{hydro}_x$ (nm)}    & $55 \pm 1$ & $52 \pm 6$ & $53\pm 20$ \\
        \footnotesize{$g$}     & $1.01\pm 0.04$ & $1.00\pm 0.02$& $1.16 \pm 0.22$ \\
        \footnotesize{$\kappa_\text{heat}$ (K$\cdot$hPa$\cdot$mW$^{-1}$)}   & $18 \pm 1$ & $12 \pm5$ & $15 \pm 10$ \\
  \hline
\end{tabular}
        \caption{Physical properties of overheated particles}
        \label{tab:Carac}
\end{table*}

\end{document}